# To be or not to be social: Foraging associations of free-ranging dogs in an urban ecosystem

Sreejani Sen Majumder[1], Anandarup Bhadra[1], Arjun Ghosh[1], Soumitra Mitra[1], Debottam Bhattacharjee[1], Jit Chatterjee[1], Anjan K. Nandi[2] and Anindita Bhadra[1*]

[1] Behaviour and Ecology Lab, Department of Biological Sciences,

Indian Institute of Science Education and Research – Kolkata, India

[2] Centre for Ecological Sciences,

Indian Institute of Science, Bangalore, India

[*]**Address for Correspondence:**

Behaviour and Ecology Lab, Department of Biological Sciences,

Indian Institute of Science Education and Research – Kolkata

P.O. BCKV Main Campus, Mohanpur,

Nadia, PIN 741252, West Bengal, INDIA

*tel.* 91-33-25873119

*fax* +91-33-25873020

*e-mail:* abhadra@iiserkol.ac.in

**Abstract**

Canids display a wide diversity of social systems, from solitary to pairs to packs, and hence they have been extensively used as model systems to understand social dynamics in natural systems. Among canids, the dog can show various levels of social organization due to the influence of humans on their lives. Though the dog is known as man's best friend and has been studied extensively as a pet, studies on the natural history, ecology and behaviour of dogs in a natural habitat are rare. Here we report results of an extensive population-level study conducted through one-time censuses in urban India to understand the ecoethology of free-ranging dogs. We built a model to test if the observed groups could have been formed through random associations while foraging. Our modeling results suggest that the dogs, like all efficient scavengers, tend to forage singly but also form random uncorrelated groups. A closer inspection of the group compositions however reveals that the foraging associations are non-random events. The tendency of adults to associate with the opposite sex in the mating season and of juveniles to stay close to adults in the non-mating season drives the population towards aggregation, in spite of the apparently random nature of the group size distribution. Hence we conclude that to be or not to be social is a matter of choice for the free-ranging dogs, and not a matter of chance.

**Keywords:** foraging association; urban ecosystem; free-ranging dogs; census; ecoethology; model

## Introduction

The canids are a fascinating family of carnivores that are highly diverse in their morphology, geographic distribution and behavioural patterns. They are the most widespread family of extant carnivora with at least one species inhabiting every continent except Antarctica, and some species spread over entire continents (Sillero-Zubiri et al. 2004). They display a wide range of social organization, from solitary living like the maned wolf (*Chrysocyon brachyurus)* to living in monogamous pairs and family units like the red fox (*Vulpes vulpes*) and arctic fox (*Alopex lagopus*) to large stable packs showing cooperative hunting and cooperative breeding behaviour like the wolves (*Canis lupus*) (Macdonald 1979; Philips et al. 2003; Sillero-Zubiri et al. 2004). Among canids, domestic dogs (*Canis lupus familiaris*) can live at diverse levels of social organization, from singly in households as pets, small groups in farms to packs in undisturbed habitats like islands (Serpell 1995), thus spanning nearly the entire range of social organization seen in canids. Though the domestic dog is known to have descended from the pack living wolves, sociality in domestic dogs has long been a matter of debate (Scott and Fuller 1965; Beck 1975; Fox et al. 1975; Kleiman and Brady 1978; Berman and Dunbar 1983; Daniels 1983; Font 1987). In fact, recent research suggests that dogs can be domesticated while wolves continuously escape attempts of domestication because of inherent differences of behaviour during early development in the two sub species (Lord 2013).

Domestic dogs that are not under direct human supervision and whose activities and movements are not restricted by human activities are termed as free-ranging dogs (Caffazo et al. 2010). Studies on populations of free-ranging dogs are widely scattered and sparse because in most developed countries dogs are not allowed to roam free on streets. In the recent years it has become quite evident that the social organization of free-ranging dogs is regulated by ecological factors that also affect other canid social systems (Macdonald and Carr 1995). In India, as in several other developing countries, dogs are commonly seen on the streets, especially in urban areas. These dogs are called strays in general, and are not under any human supervision, hence they are more aptly called free-ranging dogs (Serpell 1995). They spend their entire lives on the streets as scavengers, and though they are not owned by humans, they are dependent on humans for their sustenance (Vanak and Gompper 2009). These dogs typically have mongrel characteristics, with pointed ears, very short fur, wolf-like pointed faces and often have patch baldness in their coats (OSM Figure 1). They are an important component of the urban ecology of India, and can be found in not only cities but in towns, villages and even in forest fringes (Pal et al. 1998, Vanak and Gompper 2009). Hence they are a very good model system for studies of urban ecology and ethology and for testing models of social organization.

Urban free-ranging dogs have been studied to understand their distribution in cities, towns and fringe areas mostly in order to address the problem of strays. Jackman and Rowan (2007) has compiled several studies from developing nations in a report on the status of free-roaming dogs and methods of effective control. While some studies report that these dogs are unable to form stable social groups (Beck 1973, Berman and Dunbar 1983; Daniels 1983), others report stable social structures in the free-ranging dogs (Fox et al. 1975; Font 1987; Pal et al. 1998, Bonanni et al. 2010, Cafazzo et al. 2010). It has been argued by Beck (1973) that free-ranging dogs are asocial because the distribution of group sizes in their data matched that of a Zero-Truncated Poisson distribution (ZTP), as expected in case of a random distribution. Font (1987) made a case against this by stating that matching of the data with a ZTP distribution alone cannot be considered as proof for the dogs not forming stable social groups, and more knowledge of their behaviour is necessary to substantiate this claim. In this paper we build a model based on Poisson distribution for an expected random distribution of free-ranging dogs in space and test it with field data from dog censuses conducted in and around Kolkata, India (22°34′ N; 88°22′ E). Our results substantiate some of the arguments put forth by Font (1987). We also use the census data to build an understanding of the social tendencies of the free-ranging dogs in the urban environment.

## Methods

**(i) Sampling:** We carried out one-time censuses of free-ranging dogs at various urban localities in and around Kolkata (22° 34' N, 88° 24' E), West Bengal, India during the summer (May-June) and autumn (August-September) of 2010 and 2012. We sampled from 44 localities in the summer and from 30 localities in the autumn. The autumn months were selected for the census as this is typically the mating season for the dogs in West Bengal (Pal 2011; Sen Majumder et al, in preparation) and the summer was chosen as the non-mating season when juveniles are present. The localities were selected arbitrarily, based on convenience of sampling, and taking care that they were comparable in terms of human habitation. All localities sampled were residential or a combination of residential and business areas, because we were interested in urban dogs that live around human habitation. The absolute areas of the localities were quite variable, because the time of the census was fixed between 1600-1800 h and the observers had to cover the entire area within this time. This time was chosen as we had observed that dogs are active at this time of the day, and are typically out foraging (unpublished data), and daylight was available at this time, enabling recording of the dogs from a distance. The areas selected typically were well defined municipal blocks, or were part of a larger block bounded by arterial roads.

Each census was carried out in a single day. The day before the actual census a map of the locality was prepared with all roads and streets in the area using Google maps (http://maps.google.co.in/ ). Then the observer visited the concerned area and walked on these roads, marking the positions of the following as and when these were seen: i) waste bins ii) vats and dumps iii) food stalls (typically open roadside shanties and small shops) iv) food shops and restaurants v) markets vi) water sources like open taps, open tanks etc. The map thus prepared was used for the sampling of dogs the next day (OSM Table 1), when the observer walked along the roads and recorded any dog that was sighted, marking its approximate position on the map (Figure 1). For each dog, we recorded the time of sighting, the sex (by observing the genitalia), age class (pups, juveniles or adults, based on size and genital structures) of the dog, and whether it was single or in a group. If the dog was in a group, we also noted the group size (including the concerned dog). Groups are defined as two or more dogs that were seen to show affiliative interactions like allogrooming, nuzzling, playing, walking together, sharing food etc, or dogs that were resting peacefully within about three feet of each other. Several roads had to be walked multiple times in order to cover the entire area, but we recorded dogs on a road only the first time we walked on it, in order to avoid re-sampling. For a subset of the data we calculated the area of each locality using Google maps (http://www.daftlogic.com/projects-google-maps-area-calculator-tool.htm ) by selecting the boundaries of the locality. This could not be done for some areas as a clear area map was not available through google-maps, and the maps had been drawn manually.  StatisticXL version 1.8, STATISTICA release 7.0 and the statistical environment R (R 2008) were used for the statistical analysis.

**(ii) Modeling:** We built a model for the random distribution of the dogs in space and checked the model with our data. Let us assume $\boldsymbol{X_i}$ is the number of dogs in a group, $\boldsymbol{O_i}$ is the frequency with which $\boldsymbol{X_i}$ dogs are observed in a group, and $\boldsymbol{P(x)}$ is the probability of $\boldsymbol{x}$ dogs to be found in a group if dogs are distributed randomly over space. Then the probability distribution $\boldsymbol{P(x)}$ is expected to follow a Poisson distribution, under which, the occurrence of any dog in a group does not depend on the occurrences of the other dogs in that group, thus the numbers of dogs found in the groups are uncorrelated. Since the dogs were sampled randomly over an area and whenever a dog or a group of dogs were sighted it was noted down, so the situation of getting data of group size zero never arose, hence the 'zero' event is missing from the distribution. Therefore, we use the Zero-Truncated Poisson (ZTP) distribution which is of the form

$$P(x)=\frac{e^{-\lambda}.\lambda^{x}}{x!}\cdot\frac{1}{1-e^{-\lambda}}$$

where $\boldsymbol{\lambda}$ is the single parameter characterizing the distribution. The mean of the distribution is $\boldsymbol{\mu=\lambda/(1-e^{-\lambda})}$, and the parameter $\boldsymbol{\lambda}$ can be estimated from the equation $\boldsymbol{\mu=\langle X_i \rangle}$, thereby equating the sample mean $\boldsymbol{\langle X_i \rangle}$ with the population mean (Cohen 1960). If $\boldsymbol{E_i}$ is the expected frequency of groups containing $\boldsymbol{X_i}$ dogs, then $\boldsymbol{E_i = N.P(X_i)}$, where $\boldsymbol{N = \Sigma_i X_i O_i}$, i.e. the total frequency of the dogs.

In order to test the goodness of the fit of the data with the ZTP distribution, we used the $\boldsymbol{\chi 2}$ test. The test does not work well when expected frequencies are very small (Cochran 1952; Cochran 1954) and when testing at $\alpha$=0.05, the acceptable frequency level is 1.0 (Roscoe and Byars 1971). So, the last few minimum categories of the tail of the distribution were pooled together in order to obtain the tabulation having all expected frequencies greater than 1.0 (Cochran 1952; Zar 1999). Now if the new number of categories becomes $k$ , the degrees of freedom for the statistical test consequently becomes *v=k-1-1,* an extra *df* is lost due to the estimation of the parameter of the distribution from the data.

## Results

### i) Natural history

A total of 655 dogs were sampled from the 44 locations in the summer of which 305 were males, 331 females, and 19 were of unknown sex. In the autumn 360 dogs were sampled from the 28 locations, of which 163 were males, 189 were females and 8 were of unknown sex. Sexes could not be determined for a few pups and for a small number of adults that were found to be squatting. The sex ratio in our sample did not deviate from 1:1 in either season (t-test; t = -1.120, df = 43, p = 0.269 for the summer and t = -2.019, df = 27, p = 0.053 for the autumn). We pooled the pups (0-3 months) and juveniles (3-9 months) into the category of juveniles as the real ages of the dogs were not known, and we only had eye estimation records. The population comprised of 24 ± 19% juveniles in the summer, which was significantly higher than the proportion of juveniles (18 ± 19%) in the autumn (Mann Whitney U test, U = 880.00, df = 44, 28, p = 0.002). The total area covered in a census was quite variable as some areas were denser, with more streets and alleys than others. The mean area covered in a census was 0.09 ± 0.04 sq.km (N = 28) in the summer, with a mean dog density of 0.77 ± 0.42 dogs per acre and 0.16 ± 0.09 sq.km (N = 22) in the autumn, with a mean dog density of 0.34 ± 0.20 per acre. While the average area covered in a census was significantly higher in the autumn (Mann-Whitney U test, U = 477.0, df = 22, 28, p = 0.001), the density of dogs was significantly higher in the summer (Mann-Whitney U test, U = 518.5, df = 22, 28, p = 0.000). This is probably because there were more dogs in the summer due to the births in the winter, and by the autumn, the population had stabilized after the initial stage of high mortality of juveniles. The mean number of fixed resources present in an area, including open and closed dust bins, dumps, food stalls, restaurants and water sources was comparable between the summer and the autumn censuses (Mann-Whitney U test, U = 685.0, df = 43, 28, p = 0.334). 11 of the sampled sites did not have a market within it, but the number of dogs in areas with and without markets were comparable (Mann-Whitney U test, U = 200.5, df = 11, 32, p = 0.501). In the summer, the number of dogs in an area did not scale with the number of resources present in it (simple linear regression, $R^2 = 0.030$, $F_{1,41} = 1.276$, p = 0.265), unlike in the autumn (simple linear regression, $R^2 = 0.155$, $F_{1,26} = 4.771$, p = 0.038) (Figure 2).

**ii) Groups**

We counted the number of times $\boldsymbol{O_i}$ that the dogs were observed in a group of size $\boldsymbol{X_i}$ and named the dogs of various group sizes as solitary (size 1), paired (size 2), triad (size 3) and groups (size 4 or more). For both the seasons, we considered the proportions of dogs present in each of the groups and also in the pooled group of size four or more. 47.78 ± 18.63% of the individuals were sighted as solitary during the summer, while 40.28 ± 20.75% of the population was found to be solitary in the autumn. While there were significantly more dogs in group size 1 as compared to the other group sizes in the summer, in the autumn, the proportion of singles and pairs were comparable, and significantly higher than both the triads and higher groups (Table 1). We repeated the analysis by removing the juveniles from the data set, thereby considering only the adults, for both the seasons. We found that, by removing the juveniles from the data set, the percentage of solitary dogs changed to 57.85 ± 26.28% in the summer and 41.40 ± 21.49 in the autumn. In the summer, the removal of the juveniles from the data set caused a significant change in the proportion of solitary dogs (Wilcoxon matched pairs test, T = 144.0, N = 44, p <

0.0001) and triads (Wilcoxon matched pairs test, T = 198.0, N = 44, p = 0.017). There was no significant change in the proportions of dogs in any of the other categories, either in the summer or the autumn when the juveniles were removed (Figure 3; OSM Table 2).

Since the removal of the juveniles from the population was leading to significant changes in part of the grouping pattern, we looked at the composition of the groups in both the seasons for the entire data set. Juveniles were most often present with adults, and it was interesting to note that though 20% of the pairs were of the adult-juvenile category in the summer, there was not a single pair in this category sighted in the autumn. The proportion of pairs sighted as adult-juvenile in the autumn was significantly lower than the summer (Fisher's exact test, p = 0.0002). The adult only pairs could be male-male, female-female or male-female. The proportion of male-female pairs was 0.67 in the autumn and significantly higher than 0.32 of the summer (Fisher's exact test, p = 0.0006). The proportions of male only pairs and female only pairs did not vary in the two seasons (Fisher's exact test, p = 0.563 and 0.425 respectively; Fig. 4a). Interestingly, 47% of the juveniles were sighted as singles in the autumn, which was significantly higher than the proportion of juveniles sighted as singles (28%) in the summer (Fisher's exact test, p = 0.004). Juveniles present with males did not vary in proportion between the seasons (Fisher's exact test, p = 0.082), but the proportion of juveniles with females was higher in the summer (Fisher's exact test, p = 0.024). In both the seasons, about one third of the juveniles were sighted in juveniles-only groups, unaccompanied by any adults. Juveniles present in mixed sex groups of adults did not vary significantly in proportion between the two seasons (Fisher's exact test, p = 0.380; Fig. 4b).

### (iii) The model

The modeling exercise yielded dog distributions in the above grouping categories for the summer and autumn, both with and without the juveniles. For the summer data, the distribution of dogs in different grouping categories did not fit the Zero-Truncated Poisson distribution when we considered the entire data set ($\chi^2 = 29.528$, df = 3, $\chi^2_{0.05,3} = 7.815$), but was found to agree with the expected ZTP distribution when the juveniles were removed from the data set ($\chi^2 = 4.414$, df = 2, $\chi^2_{0.05,2} = 5.991$). When we carried out similar operations on the autumn data, the distribution fitted well into the ZTP distribution for both the whole data set ($\chi^2 = 3.470$, df = 3, $\chi^2_{0.05,3} = 7.815$) and the one with the juveniles removed ($\chi^2 = 2.064$, df = 3, $\chi^2_{0.05,3} = 7.815$). Thus the dogs appeared to be randomly distributed in space at the time of foraging, unless they were with juveniles.

## Discussion

Free-ranging dogs have been reported to have a male biased sex ratio in the USA and Europe (Beck 1973, Daniels 1983, Daniels and Bekoff 1989). Beck (1973) suggested that males are taken more often as pets, and since most urban feral dogs are those that have been abandoned or have run away from domestication, the sex ratio in the feral population is biased. Moreover, females might be killed in order to reduce breeding, or may be selectively abandoned as pups. However, these results pertain to "feral" dogs with an immediate history of domestication, and could be quite different behaviourally from the Indian free-ranging dogs. Pal (2008) reported a male biased sex ratio of the free-ranging dogs in Katwa, West Bengal, India, both at birth and among the adult population from a study conducted on six bitches and their pups. However, in our population level study conducted over 71 localities, the sex ratio did not deviate significantly from 1:1 in a total sample size of 1015 dogs. It is possible that male pups are indeed adopted as pets preferentially, and this leads to the evening out of the sexes in the population, in spite of the male biased sex ratio at birth.

Dogs are known to breed twice a year (Morris 1987), though an individual bitch usually comes into heat once every year. The free-ranging dogs in West Bengal primarily mate in the autumn (Pal 2011) but we have also observed some matings in the late spring (April-May, unpublished data). The gestation period in dogs is approximately two months (Morris 1987), and thus when they mate in the autumn, the pups are born in the winter, resulting in a large number of juveniles in the population during the summer. The juveniles are typically in the post-weaning phase (3-9 months), and are not restricted to the shelters. Since this study was conducted in May-June and August-September, it was unlikely that pups born due to matings in the spring would have been present in the summer data. In the autumn, such pups, if any, would also be close to the weaning stage of 10 weeks (Paul et al, under review), and would not be restricted to the shelters (Pal 2008). Hence at the time of our census, we were likely to find them on the streets with the adults, and chances of missing them were low.

We were primarily interested in studying the distribution of the dogs during their active period, i.e., when they are likely to forage. The urban free-ranging dogs are scavengers living in a highly competitive environment, where resources can be quite diffused and unstable. It is known that the spatial distribution and social organization of animals are affected by the distribution of key resources (Macdonald 1983, Johnson et al 2003). In our study, the dog numbers in an area were not dependent on the number of available resources in the summer, but scaled with the number of resources in the autumn. This difference in the relationship between dog numbers and resource availability between the two seasons could be attributed to the higher proportion of juveniles in the summer and the fact that reproduction in an unstable environment is not expected to scale with resource availability. However, since the resources that the dogs depend on range from large dumping sites to friendly humans, number alone is perhaps not a very good estimate of resource abundance and richness of an area. Currently we are carrying out detailed observations of dog behaviour at feeding sites to better understand the pattern of resource utilization by the free-ranging dogs and how this affects their social behaviour. Such data, in combination with data from censuses carried out over large areas would not only provide an insight into the resource utilization pattern and social organization of the free-ranging dogs, but will also allow us to use the dogs as a model system to test theories like the resource dispersion hypothesis (Macdonald 1983, Johnson et al 2002) with field data.

Dogs are known to have descended from wolves that live and hunt in packs (Mech 1970), and have been shown to be social in several studies (Font 1987; Pal et al. 1998, Cafazzo et al 2010). In our model, the distribution of the dogs in space fitted the ZTP distribution for the autumn data when the entire data set was considered, as well as when the juveniles were removed from the population. For the summer the distribution fitted the ZTP only when the juveniles were removed from the data set. These results suggest that the dogs form random uncorrelated groups at the time of foraging, as reported earlier by Beck (1973), so that the probability of a new dog joining a group is independent of the presence of the existing dogs in that group. An alternative to this could only be one of the following two situations. The distribution can be biased towards uniformity, such that the occurrence of one dog in a group impedes that of the second dog in that group. In this case we would obtain repulsed, and thus, negatively correlated groups of dogs and thereby could call them asocial. The second alternative is that the population is biased towards aggregation or clustering. Here the probability of the occurrence of the first dog in a group enhances the probability of occurrence of the second one in that group, therefore developing a positive correlation among the dogs. The second case is indeed what is observed in the summer data when juveniles are present – they prefer to stay with the adults, thus making the distribution contagiously non-random.

On closer examination of the group compositions, we realized that though the global nature of the distribution appeared to be random, the composition of the groups were not so random after all. There was a clear preference for adult male-female pairs in the mating season and a preference for foraging singly in the non-mating season, suggesting that the dogs try to avoid competition over foraging, but also may choose to forage in association with preferred partners in certain contexts, like mating and parental care. This is borne out by the fact that though nearly half of the dogs were sighted as solitary, this fraction was not constant in the two seasons. The proportion of solitary dogs was higher than all the other categories in the summer, but in the autumn this proportion, though still nearly 40%, was comparable to that of the pairs. Hence during the mating season the dogs tended to be together more often than during the non-mating season, even at the cost of facing competition over food. This intriguing pattern in group dynamics suggests that the distribution of resources and competition over them might be playing key roles in determining the social interactions that shape groups in the free-ranging dogs. We should remember that the study was conducted during the time of day when the dogs are usually active, and the distribution studied here refers only to the associations during foraging, which might be very different from the grouping at the time of resting or territory defense, as suggested by Font (1987). In fact, our observations suggest that the dogs tend to defend territories in groups which they also adhere to during resting, but tend to forage in smaller subgroups or singly (Das and Bhadra, in preparation). Hence we can be all the more certain that the associations seen during foraging are a result of the choices of the individuals, and not random associations of unfamiliar dogs, as the case might be if the dogs are indeed randomly distributed in space. We confirm through our model that the distribution of the free-ranging dogs in space during foraging has a globally random nature, but local associations are indeed an outcome of individual preferences to accept competition and yet stay in a group or to be solitary to avoid competition and thereby also give up the advantages of being social.

**Acknowledgements**

This work was supported by grants from the Council for Scientific and Industrial Research, India and the Indian National Science Academy to AB, and by IISER-Kolkata. AKN carried out the modeling and all the remaining authors conducted spot censuses in different locations and times, and appear in the list of authors according to the volume of work done in the field. AB supervised the work and co-wrote the paper with AKN. AKN wishes to thank Dr. Kunal Bhattacharya, Birla Institute of Technology and Science, Pilani, India for his valuable feedback on the modeling part. The authors are grateful to Prof. Raghavendra Gadagkar, Indian Institute of Science, Bangalore, India and three anonymous referees for their inputs on earlier versions of this manuscript.

| | Summer (N = 44) | | Autumn (N = 28) | |
|---|---|---|---|---|
| **Comparisons** | **T** | **p** | **T** | **p** |
| **Solitary vs Paired** | 124 | < 0.0001 | 163.50 | 0.4740 |
| **Solitary vs Triad** | 84 | < 0.0001 | 81.00 | 0.0070 |
| **Solitary vs Grouped** | 47.50 | < 0.0001 | 11.00 | < 0.0001 |
| **Paired vs Triad** | 366.50 | 0.1510 | 106.50 | 0.0270 |
| **Paired vs Grouped** | 224.50 | 0.0020 | 10.00 | < 0.0001 |
| **Triad vs Grouped** | 304.50 | 0.0740 | 64.00 | 0.0030 |

**Table 1:** Summary of the comparisons between the four kinds of group sizes in the two seasons using Wilcoxon matched-pairs test. All comparisons are within a season between group sizes.

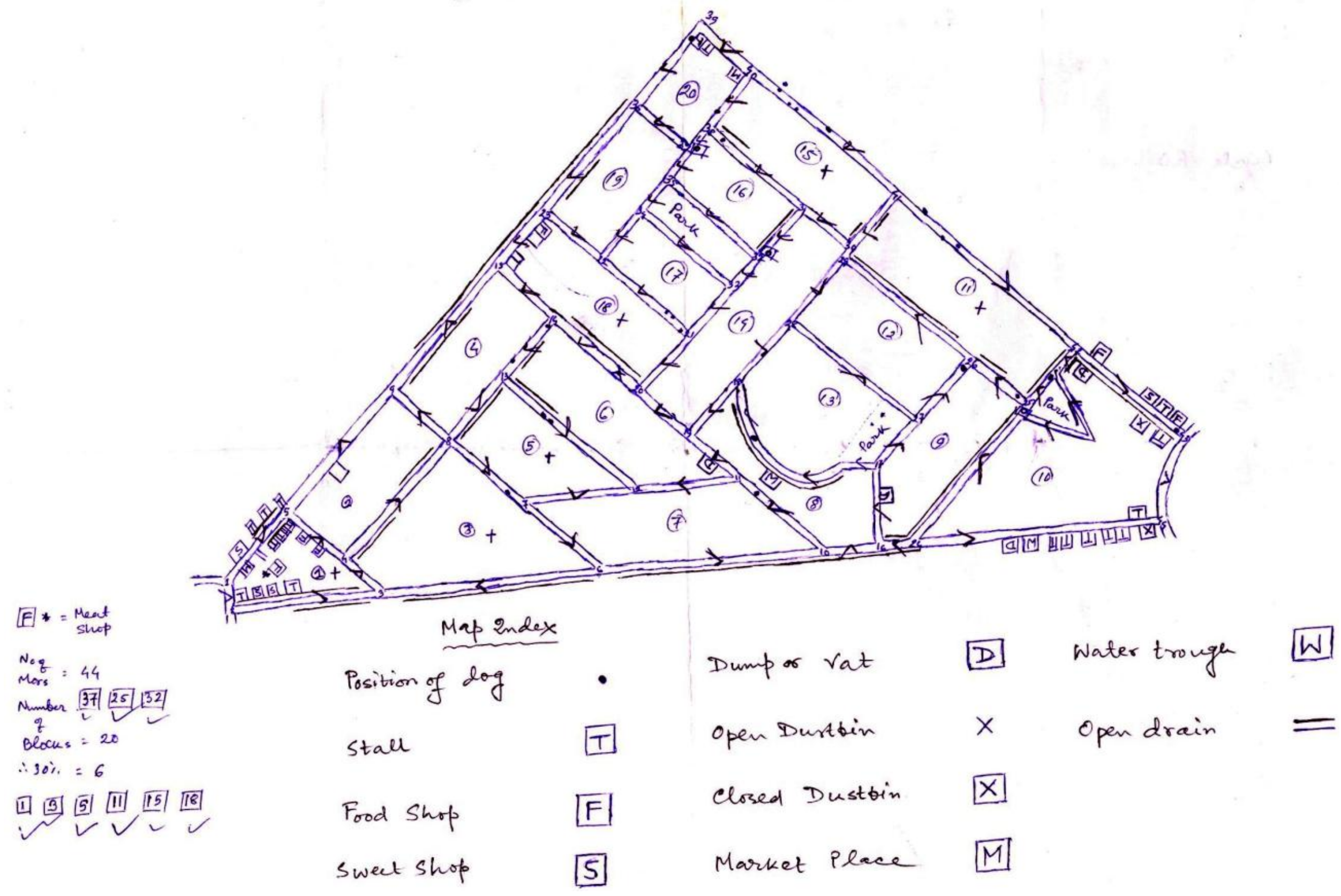

**Figure 1:** An example of a map used for sampling (part of B-6 block of Kalyani). The arrows show the path followed for conducting the census, and various resources are marked using the index given at the bottom of the map.

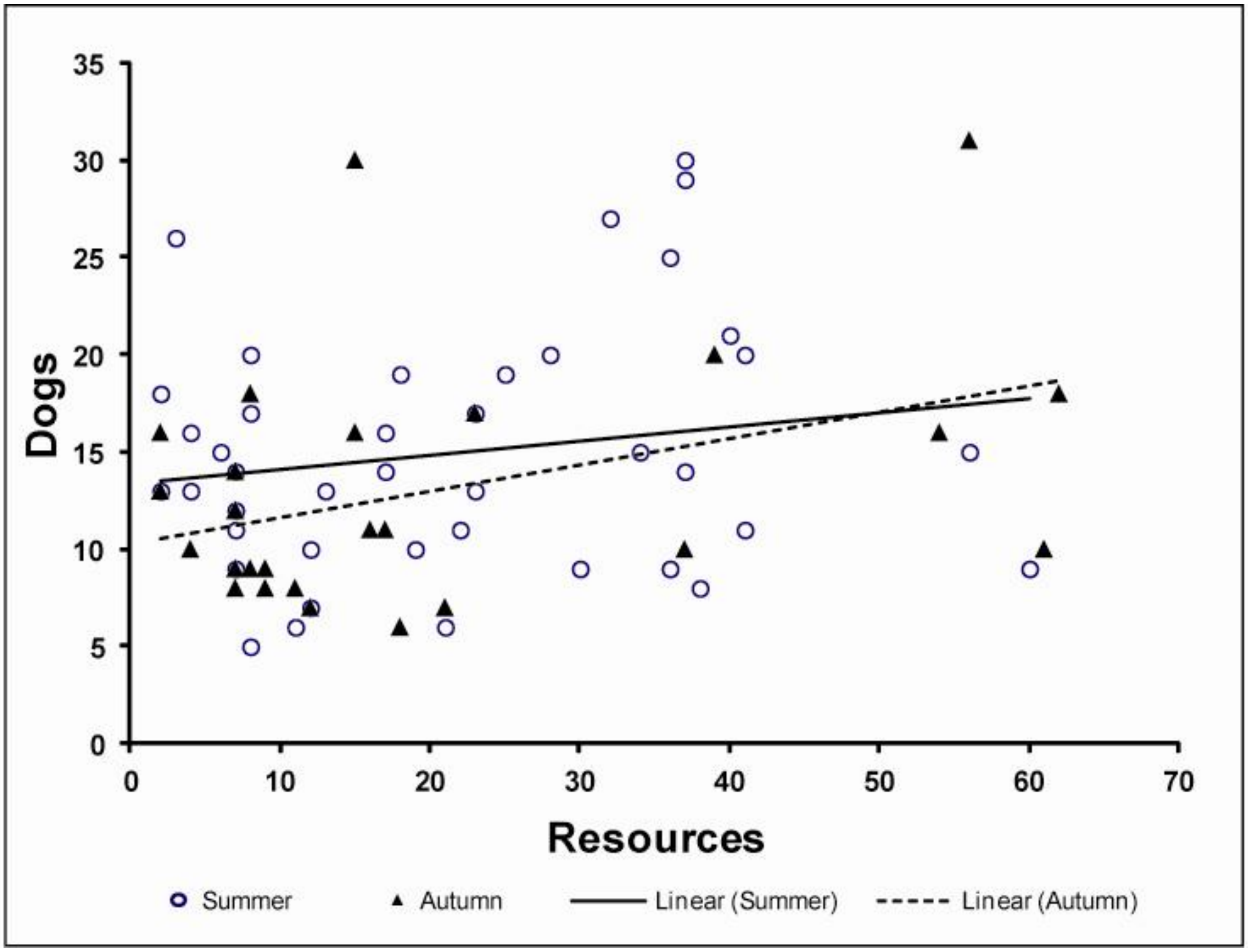

**Figure 2:** A scatter plot showing the number of resources and the number of dogs recorded in each census in both the seasons (summer: circles and autumn: triangles). The linear regression lines for both seasons are also given.

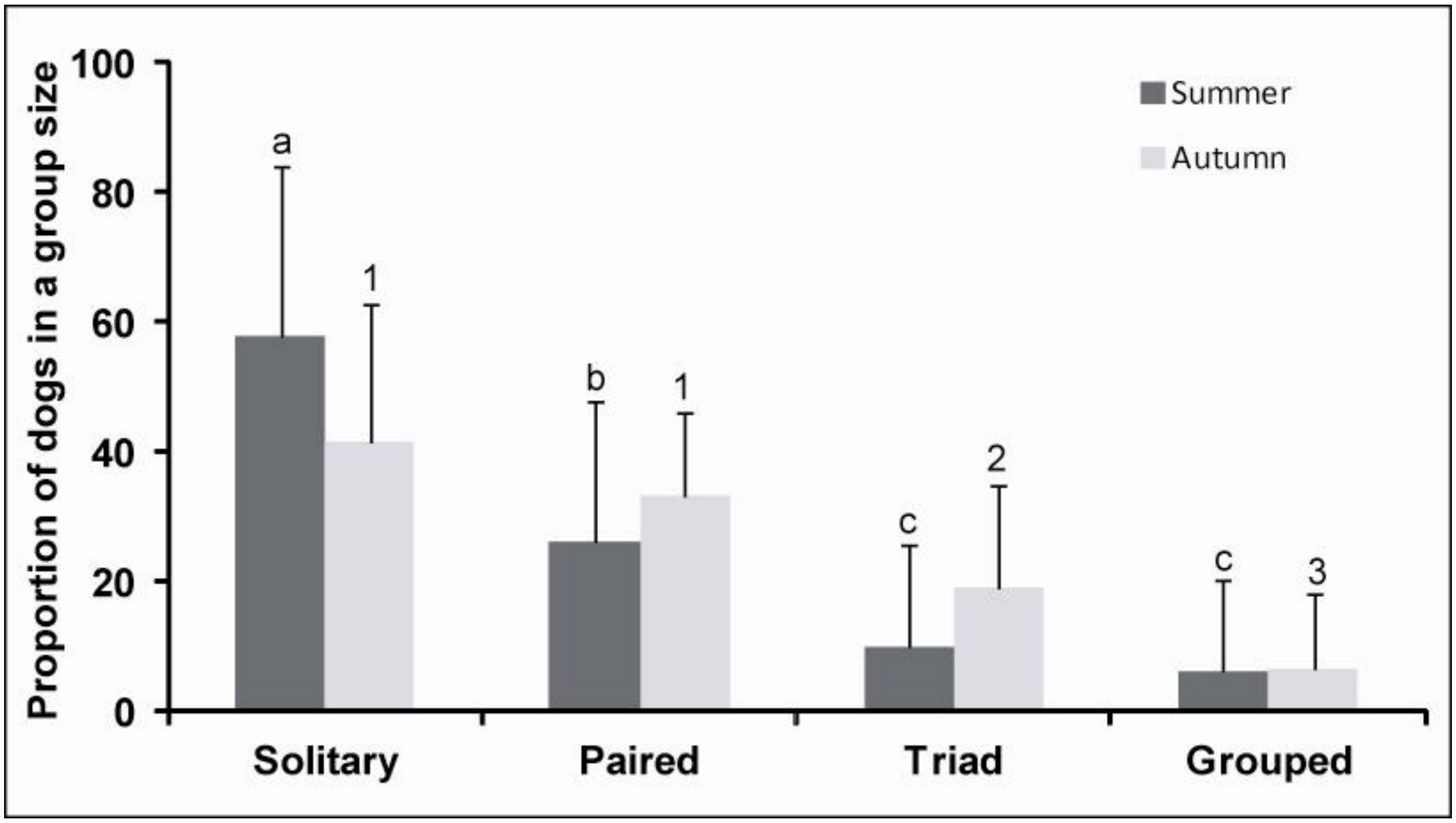

**Figure 3:** Mean and S.D. of the proportion of adult dogs found as solilitary, in pairs, triads and in groups of 4 or more in the two seasons. Comparisons are between categories, within a season, using Wilcoxon matched-pairs test (significance at $p < 0.05$).

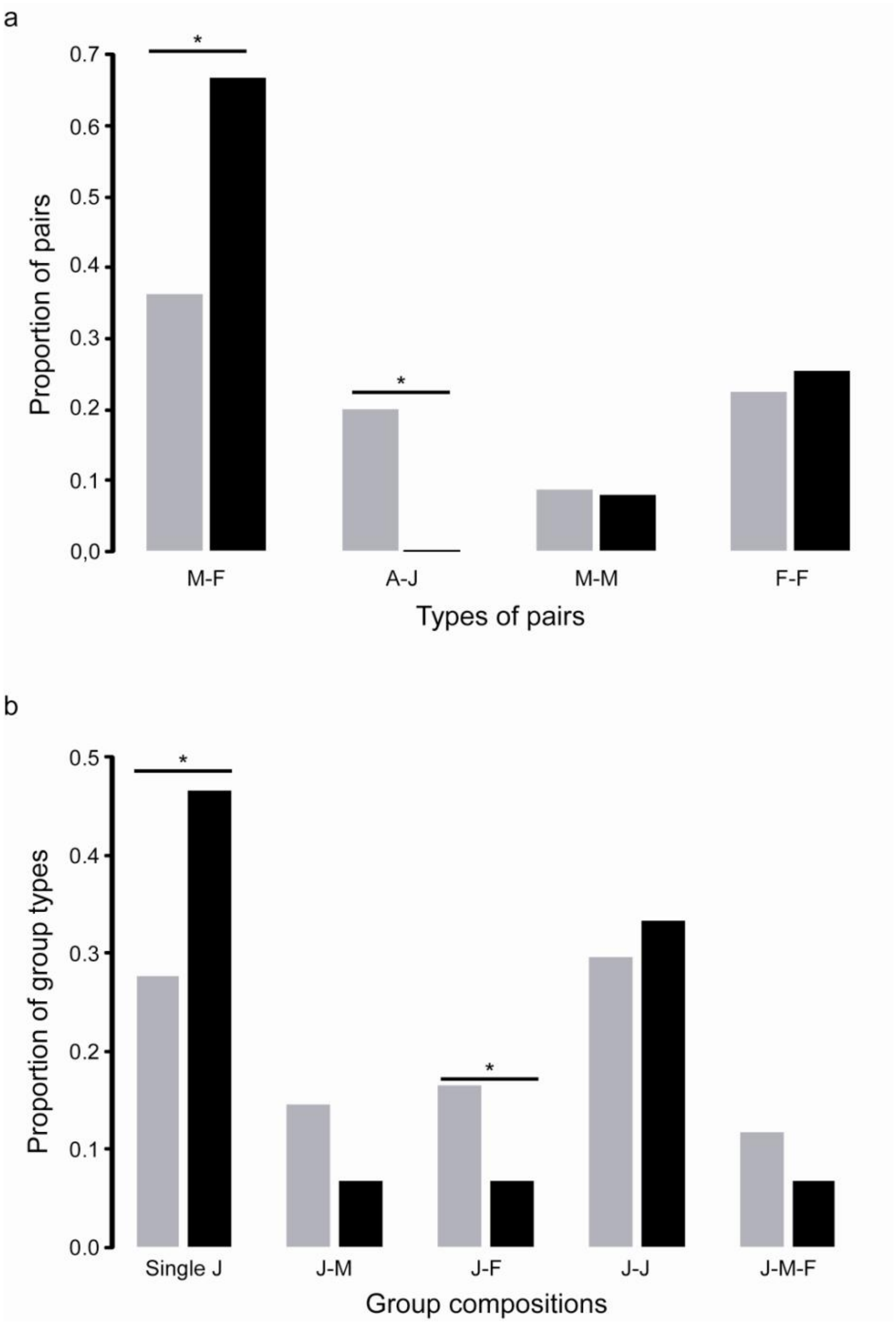

**Figure 4:** (a) The distribution of the proportions of the different kinds of pairs observed in the summer (gray bars) and autumn (black bars). (b) The distribution of the proportions of different group compositions (all group sizes other than single combined together) in which the juveniles are distributed in the summer (gray bars) and autumn (black bars). A: adults, J: juveniles, M: males, F: females. Statistically significant differences are indicated by asterisk (*).

**Supplementary Material**

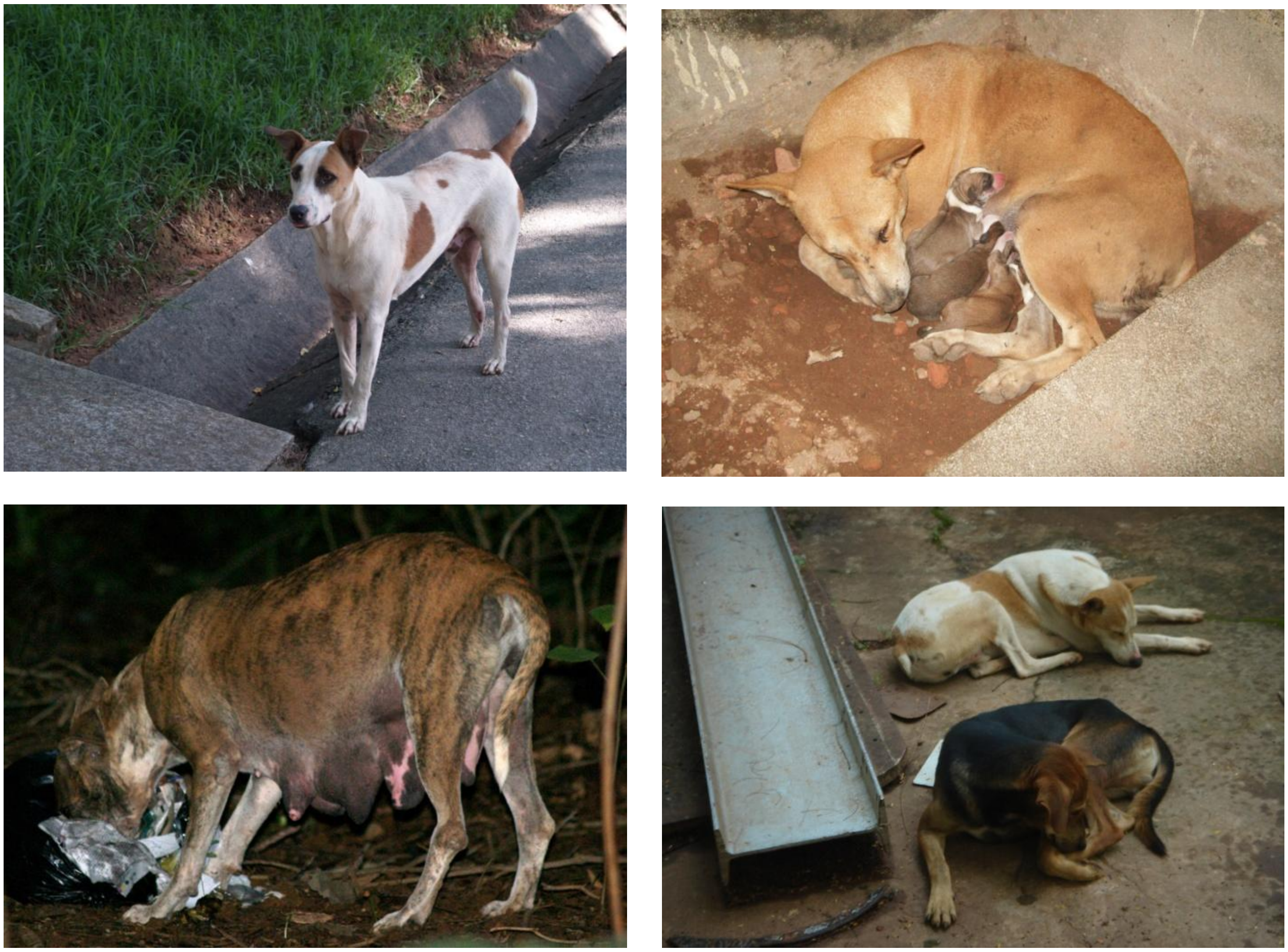

**SM Figure 1:** Free-ranging dogs in India have pointed ears, short fur and pie baldness. They live among humans, spending most of the time on streets. They depend on garbage and human generosity for their sustenance, and are rarely seen to hunt.

**Methods**

| Sl No | Observer | Date | Time | Group size | Age Class | Sex |
|---|---|---|---|---|---|---|
| C-0022/001 | SOUMITRA | 27.06.10 | 16:26 | 1 | A | M |
| C-0022/002 | SOUMITRA | 27.06.10 | 16:29 | 1 | A | F |
| C-0022/003 | SOUMITRA | 27.06.10 | 16:40 | 1 | A | M |
| C-0022/004 | SOUMITRA | 27.06.10 | 16:42 | 1 | A | F |
| C-0022/005 | SOUMITRA | 27.06.10 | 16:50 | 1 | A | M |
| C-0022/006 | SOUMITRA | 27.06.10 | 16:54 | 1 | A | M |
| C-0022/007 | SOUMITRA | 27.06.10 | 16:17 | 1 | A | F |
| C-0022/008 | SOUMITRA | 27.06.10 | 16:39 | 2 | A | M |
| C-0022/009 | SOUMITRA | 27.06.10 | 16:39 | 2 | A | M |
| C-0022/001 | SOUMITRA | 27.06.10 | 16:26 | 1 | A | M |
| C-0022/002 | SOUMITRA | 27.06.10 | 16:29 | 1 | A | F |
| C-0022/003 | SOUMITRA | 27.06.10 | 16:40 | 1 | A | M |
| C-0022/004 | SOUMITRA | 27.06.10 | 16:42 | 1 | A | F |

**SM Table 1:** Sample data from one census in Bankura conducted in the summer of 2010.

**Results**

| | Summer (N = 44) | | Autumn (N = 28) | |
|---|---|---|---|---|
| **Comparisons** | **T** | **p** | **T** | **p** |
| **Solitary** | 144.00 | < 0.0001 | 26.00 | 0.250 |
| **Paired** | 349.50 | 0.2740 | 27.00 | 0.313 |
| **Triad** | 198.00 | 0.0170 | 51.00 | 0.750 |
| **Grouped** | 135.00 | 0.0930 | 27.00 | 1.00 |

**SM Table 2:** Comparisons between the adult-only data set and the entire data set in the four group sizes in the two seasons. All comparisons are within a group size in a season using Wilcoxon matched-pairs test.